\documentclass{emulateapj}

\slugcomment{ApJ received 2013 December 5; aceppted 2014 June 13}
\shorttitle{Techniques for detecting intergalactic magnetic fields}
\shortauthors{Akahori, Gaensler, Ryu}

\begin{document}
\title{Statistical Techniques for Detecting the Intergalactic Magnetic Field from Large Samples of Extragalactic Faraday Rotation Data}
\author{
Takuya Akahori$^1$, B. M. Gaensler$^1$, and Dongsu Ryu$^2$
}
\affil{$^1$Sydney Institute for Astronomy, School of Physics, The University of Sydney, NSW 2006, Australia: \\
akahori@physics.usyd.edu.au, bryan.gaensler@sydney.edu.au \\
$^2$Department of Physics, UNIST, Ulsan 689-798, Korea: ryu@sirius.unist.ac.kr
}

\begin{abstract}
Rotation measure (RM) grids of extragalactic radio sources have been widely used for studying cosmic magnetism. But their potential for exploring the intergalactic magnetic field (IGMF) in filaments of galaxies is unclear, since other Faraday-rotation media such as the radio source itself, intervening galaxies, and the interstellar medium of our Galaxy are all significant contributors. We study statistical techniques for discriminating the Faraday rotation of filaments from other sources of Faraday rotation in future large-scale surveys of radio polarization. We consider a $30^\circ\times 30^\circ$ field-of-view toward the south Galactic pole, while varying the number of sources detected in both present and future observations. We select sources located at high redshifts and toward which depolarization and optical absorption systems are not observed, so as to reduce the RM contributions from the sources and intervening galaxies. It is found that a high-pass filter can satisfactorily reduce the RM contribution from the Galaxy, since the angular scale of this component toward high Galactic latitudes would be much larger than that expected for the IGMF. Present observations do not yet provide a sufficient source density to be able to estimate the RM of filaments. However, from the proposed approach with forthcoming surveys, we predict significant residuals of RM that should be ascribable to filaments. The predicted structure of the IGMF down to scales of $0.1^\circ$ should be observable with data from the SKA, if we achieve selections of sources toward which sightlines do not contain intervening galaxies and RM errors are less than a few rad m$^{-2}$.

\end{abstract}

\section{Introduction}
\label{section1}

The intergalactic medium (IGM) in the cosmic web of filaments and clusters of galaxies is thought to be permeated with an intergalactic magnetic field (IGMF). Understanding the properties of the IGMF is essential for elucidating radiative processes and particle acceleration in the cosmic web \citep[see,][for reviews]{gbf04,ryu12,fer12}. Faraday rotation measures (RMs) of polarized extragalactic radio sources is a promising approach for studying the IGMF. The RM from a background source located at a redshift $z_{\rm s}$ seen by an observer at $z=0$ can be written as
\begin{equation}\label{eq:RM}
{\rm RM(z_{\rm s})}\approx 812 \int_{z_{\rm s}}^{0}
\frac{n_{\rm e}(z) B_\parallel(z)}{(1+z)^{2}}
\frac{dl(z)}{dz}dz
~{\rm rad~m^{-2}}~,
\end{equation}
where $n_{\rm e}(z)$ the electron density at a redshift $z$ in units of ${\rm cm^{-3}}$, $B_\parallel(z)$ is the line-of-sight (LOS) component of the magnetic field at $z$ in $\mu$G, and $dl(z)$ is a line element along the LOS at $z$ in kpc. 

The potential of RM grids for studying extragalactic magnetic fields has been demonstrated for galaxies and galaxy clusters \citep[e.g.,][]{ckb01,gae05}. But that for filaments of galaxies is yet to be established, since other sources of Faraday rotation along the LOS are not negligible compared to the expected IGMF RM of $\sim 1-10~{\rm rad~m^{-2}}$ through filaments \citep{ar10,ar11}. For example, RMs of a few to several hundreds of ${\rm rad~m^{-2}}$ are usually associated with both the background extragalactic radio sources themselves \citep[e.g.,][]{osu12} and with the Galactic magnetic field (GMF) in our own Milky Way \citep[e.g.,][]{opp12}.  Errors in RM observations are $\lesssim 10~{\rm rad~m^{-2}}$ \citep[e.g.,][]{mao10,sts11}, including RMs of $\sim$a few ${\rm rad~m^{-2}}$ due to the Earth's ionosphere \citep[e.g.,][]{sb13}. Faraday rotation in intervening galaxies may also occur along the LOS \citep[e.g.,][]{kro08,ber08,ber12}. Therefore, we need techniques for separating these other sources of Faraday rotation from Faraday rotation through filaments and large-scale structure.

The separation can be partly possible by considering spatial correlation and dependence of RMs. For example, a high-pass filter can be employed to remove the Galactic contribution \citep[the component remaining after filtering is often called the residual RM or RRM, see][and references therein]{ham12}. \citet{sch10} examined the latitude dependence of RM in the VLA data of \cite{tss09}, and estimated that the standard deviations of RMs for Galactic and extragalactic contributions are $\sim 6.8 \pm 0.1 (8.4 \pm 0.1)~{\rm rad~m^{-2}}$ and $\sim 6.5 \pm 0.1 (5.9 \pm 0.2)~{\rm rad~m^{-2}}$ for the northern (southern) hemisphere, respectively. An analysis of the structure function (SF) of RM is also insightful. \cite{arkg13} simulated the Galactic RM toward high Galactic latitudes, and concluded that the amplitude and slope of observed SFs \citep{mao10,sts11} are both hard to explain if only the Galactic contribution is present. They concluded that there must be additional, small-scale ($\lesssim 1^\circ$) Faraday rotation in the data, possibly corresponding to an extragalactic or intergalactic component. Alternatively, the two-point correlation of RMs \citep{kol98} and the cross-correlation between RMs and galaxies \citep{xu06,sta10} can also provide constraints on the structure of the IGMF.

Another powerful discriminant is the correlation between RM (or RRM) and $z_{\rm s}$ \citep[e.g.,][]{kro08,ber12,ham12}. \cite{ham12} suggested that $10-15~{\rm rad~m^{-2}}$ of the RRM signal seen in the VLA data could be extragalactic contributions that must originate between the polarized radio sources and our Galaxy.

The above statistical techniques can be improved with more data. Indeed, a very large number of extragalactic RMs will be detected in future observations with the Square Kilometre Array (SKA) and its precursors such as the Australian SKA Pathfinder (ASKAP). Therefore, in this paper we examine statistical techniques and clarify the potential of RM grids for studying the IGMF in filaments of galaxies. We investigate the ways in which the statistics will improve in future observations with higher RM sky densities, and we consider the corresponding constraints that can then be obtained on the IGMF using these data.  Our approach is to create mock RM maps, and then use them to search for the statistical signature of the RM due to the IGMF. In Section 2, we describe our model. Our calculations are explained in Section 3. The results are shown in Section 4, and the discussion and conclusion follow in Sections 5 and 6, respectively.

\section{Model}
\label{section2}

\begin{table*}[tp]
\begin{center}
\caption{
RM components assumed as the most probable scenario in this study.\label{table1}
}
\begin{tabular}{ccrrl}
\tableline\tableline
Tag & Component & Average, $\mu$ & Deviation, $\sigma$ & Reference \\
& & ${\rm rad~m^{-2}}$ & ${\rm rad~m^{-2}}$ & \\
\tableline
INT & Intrinsic to source & 0.0 & 3.1 (10.0 at $z=0$) & \cite{ham12} \\
IGM & Intergalactic medium& 0.0 & 7.2 & \cite{ar11} \\
EXG & External galaxies & - & - & \cite{ber12} \\
ISM & Interstellar medium & +7.6 & 8.9 & \cite{arkg13} \\
ERR & Observational error & 0.0 & 1.0 & \cite{sb13}\\
COM & Combination of all & +7.6 & 11.7 & \\
\tableline
\end{tabular}
\end{center}
\end{table*}

We consider multiple RM components along the LOS: the intrinsic RM associated with a polarized extragalactic radio source (hereafter labeled INT), the RM of the IGM (IGM), the RM of any intervening external galaxies (EXG), the RM of the ionized interstellar medium in our Galaxy (ISM), and the RM caused by possible errors in the observations (ERR). Following our previous study \citep{arkg13}, we consider a field-of-view (FOV) toward high Galactic latitudes, where the Galactic contribution to RM is smallest. 

The observed RM is the combination (COM) of the above components. An average and a standard deviation of RM, $\mu$ and $\sigma$ respectively, for sources within a given FOV can be written as
\begin{equation}\label{eq:mu}
\mu_{\rm COM} = \mu_{\rm INT} + \mu_{\rm IGM} + \mu_{\rm EXG} + \mu_{\rm ISM} + \mu_{\rm ERR},
\end {equation}
\begin{equation}\label{eq:sigma}
\sigma_{\rm COM}^2 = \sigma_{\rm INT}^2 + \sigma_{\rm IGM}^2 + \sigma_{\rm EXG}^2 + \sigma_{\rm ISM}^2 + \sigma_{\rm ERR}^2,
\end {equation}
where each component is measured in the observer's frame. Below, we present simple scenarios for each of the RM components (Table 1). The results for other scenarios are shown in Section~\ref{section5}.

\subsection{RM Associated with the Source (INT)}
\label{subsection2.1}

\begin{figure}[tp]
\includegraphics[width=\linewidth]{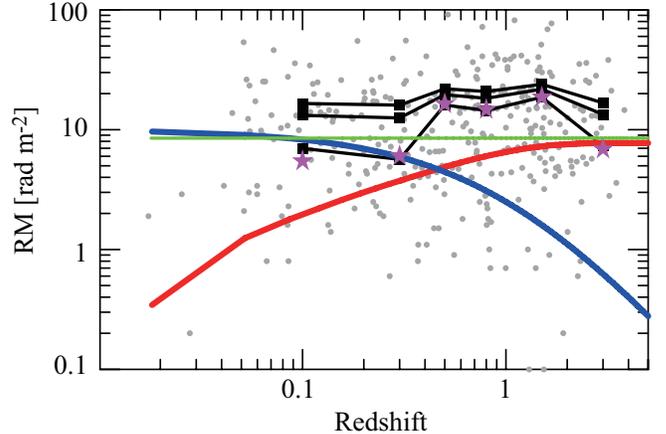}
\caption{
RM as a function of redshift. Gray filled circles show the observed RMs and redshifts of 317 sources located at $|b|>75^\circ$ \citep{ham12}. Black lines with filled squares, from top to bottom, indicate the standard deviations of error-subtracted RMs, $\sigma_{\rm RM}^{*}=(\sigma_{\rm RM}^2-\sigma_{\rm ERR}^2)^{1/2}$, for $\sigma_{\rm ERR}=$ 0, 10, and 15 ${\rm rad~m^{-2}}$, respectively, binned with redshift ranges, $z=0-0.2$ (71 sources), $0.2-0.4$ (50), $0.4-0.6$ (31), $0.6-1.0$ (55), $1.0-2.0$ (77), and $2.0-4.0$ (33). The blue line shows $\sigma_{\rm INT}=\sigma_{{\rm INT},0}(1+z)^{-2}$ with $\sigma_{{\rm INT},0}=10$ ${\rm rad~m^{-2}}$. The red line shows $\sigma_{\rm IGM}$ in the TS0 run \citep{ar11}. The green line shows $\sigma_{\rm ISM}=8.4$ ${\rm rad~m^{-2}}$ \citep{sch10}. Magenta stars show the residual, $\sigma_{\rm EXG}=(\sigma_{\rm RM}^2-\sigma_{\rm INT}^2-\sigma_{\rm IGM}^2-\sigma_{\rm ISM}^2-\sigma_{\rm ERR}^2)^{1/2}$ with $\sigma_{\rm ERR}= 10.0$ ${\rm rad~m^{-2}}$. \label{f1}
}
\end{figure}

\placefigure{f1}

We assume that intrinsic RMs of extragalactic polarized sources are spatially uncorrelated and follow a random Gaussian distribution for a given redshift bin, $[z,z+dz]$, with an average $\mu_{{\rm INT},z}=0$ ${\rm rad~m^{-2}}$, and a standard deviation $\sigma_{{\rm INT},z}$. Here the subscript $z$ means the quantity in the source frame. For the redshift distribution of sources, we employ a model based on observations \citep{wil08,ar11}

The redshift dependence that should be adopted for $\sigma_{{\rm INT},z}$ is not clear. Cosmological simulations have suggested that the rest-frame intrinsic RM associated with a starburst galaxy increases with redshift, since the associated density and magnetic-field strength both increase \citep{bec12}. On the other hand, beam depolarization which reduces the magnitude of the observed RM \citep[e.g.,][]{sok98} can take place for high-redshift sources for which angular scales of magnetic fields become much smaller than the beam size. Figure \ref{f1} shows the RMs of 317 sources at Galactic latitudes $|b|>75^\circ$, where the RMs are taken from \cite{tss09} and the corresponding redshifts are from \cite{ham12}. This plot indicates that observed RMs are not correlated with redshift.

Whatever the redshift dependence of $\sigma_{{\rm INT},z}$, the observed RM will be a factor of $1/(1+z)^2$ times smaller than the value in the frame in which the Faraday rotation occurs. This dilution with redshift is not seen in observed RMs either. Therefore, unless $\sigma_{{\rm INT},z}$ increases with redshift, in the data there must be other contributors whose RMs increase with redshift. We expect that the contributor is the IGMF (the red line in Figure~\ref{f1}, see Section 2.2). We adopt this scenario; significant RM of the IGMF exists in the data, and consider a simple case that $\sigma_{{\rm INT},z}$ does not evolve with redshift, i.e. $\sigma_{{\rm INT},z}=\sigma_{{\rm INT},0}$. The RM at the observer is thus $\sigma_{\rm INT}=\sigma_{{\rm INT},0}(1+z)^{-2}$ as shown by the blue line in Figure~\ref{f1}.

We estimate $\sigma_{{\rm INT},0}$ from observed RMs as follows. The standard deviation of the 317 RMs is 19.9 ${\rm rad~m^{-2}}$, which is larger than the $\sim 9$ ${\rm rad~m^{-2}}$ observed for the WSRT and ATCA sources with $|b|>75^\circ$ \citep{mao10}. The data include errors, so that the discrepancy may be attributed to the difference in noise power between observations. This implies that $\sigma_{\rm ERR}\sim 18$ ${\rm rad~m^{-2}}$ for the VLA RMs, although such a value is larger than the standard deviations in the two lowest and one highest redshift bins (black filled squares in Figure~\ref{f1}). If we adopt $\sigma_{\rm ERR}=10-15$ ${\rm rad~m^{-2}}$ \citep{sch10,sts11}, the standard deviation of error-subtracted RMs, $\sigma_{\rm RM}^*$, for the lowest redshift bin is 7.0--13.2 ${\rm rad~m^{-2}}$. We could ascribe this to the root of $\sigma_{{\rm INT},0}^2+\sigma_{\rm ISM}^2$, since $\sigma_{\rm IGM}$ and $\sigma_{\rm EXG}$ could be small for their short path lengths. Considering a Galactic contribution of $6.8-8.4$ ${\rm rad~m^{-2}}$ \citep{sch10}, we adopt $\sigma_{{\rm INT},0}=10$ ${\rm rad~m^{-2}}$ and $\mu_{\rm INT}=0$ ${\rm rad~m^{-2}}$ in this paper. The corresponding value of $\sigma_{{\rm INT}}$ for all sources is $\sim 3.1~{\rm rad~m^{-2}}$ in the observer's frame.

\subsection{RM Associated with IGMF (IGM)}
\label{subsection2.2}

We adopt a model for the intergalactic RM due to the IGMF in filaments of galaxies as calculated by \cite{ar11}. A model IGMF is based on a turbulent dynamo \citep{rkcd08}. The average strength of the IGMF in fiaments is $\langle B\rangle \sim O(10)$ nG or $\langle \rho B\rangle/\langle \rho\rangle \sim O(100)$ nG at $z=0$, where $B$ is the IGMF strength and $\rho$ is the IGM density. The characteristic scale of the IGMF in filaments is several hundreds of ${\rm kpc}$.

We integrate the IGM RM from a source located at $z_{\rm s}$ to an observer located at the center of a group of galaxies at $z=0$. The integration contains the contribution from the Local Group, whose RM is small, $\lesssim O(10^{-1})$ ${\rm rad~m^{-2}}$. We choose the TS0 run of \cite{ar11}, in which LOSs passing through galaxy clusters are excluded. The clusters are identified from the criteria of X-ray surface temperature and X-ray surface brightness, so as to mimic the detection limit of future X-ray facilities. Note that the current X-ray detection limit is already sufficient to substantially exclude sources located behind galaxy clusters from RM grids \citep[see][]{ar11}. In the TS0 run, $\mu_{\rm IGM}$ is set to be $0$ ${\rm rad~m^{-2}}$, and $\sigma_{\rm IGM}$ overtakes $\sigma_{\rm INT}$ at $z\gtrsim 0.5$ and reaches $\sigma_{\rm IGM}\sim 7.2$ ${\rm rad~m^{-2}}$ for filaments up to $z=5$, shown by the red line in Figure~\ref{f1}. See \cite{ar11} for details.

\subsection{RM Associated with GMF (ISM)}
\label{subsection2.3}

We adopt a model for the Galactic RM due to the GMF toward the South Galactic pole as calculated by \cite{arkg13}. We choose their ADPS30 run as a representative model. In the ADPS30 run, regular components are modelled using the electron density model of \cite{cl02} and using the GMF models consisting of an axi-symmetric spiral field and a halo toroidal field \citep{sun08} plus a dipole poloidal field that produces a vertical field near the Sun \citep{gkss10}. Random components of the density and magnetic fields are modelled using MHD turbulence simulations \citep{krjh99} with an rms flow speed of $30~{\rm km/s}$ and a driving scale of $250~{\rm pc}$ \citep{hil08}. The strength of the regular magnetic field is a few $\mu$G near the disk and smaller at higher altitudes. The strength of the turbulent magnetic field is at most a few $\mu{\rm G}$ and mostly $\la 1~\mu{\rm G}$. See \cite{arkg13} for details.

For a $30^\circ \times 30^\circ$ FOV toward the South Galactic pole, the model gives $\sigma_{\rm ISM}\sim 5$ ${\rm rad~m^{-2}}$ which is smaller than the observed estimate of $8.4\pm 0.1$ ${\rm rad~m^{-2}}$ \citep[][the green line in Figure~\ref{f1}]{sch10}. To increase the simulated $\sigma_{\rm ISM}$, we introduce a constant multiplicative factor into the calculation of RM, i.e. we considered somewhat larger densities and magnetic fields than originally simulated, to obtain $\sigma_{\rm ISM}\sim 8.9$ ${\rm rad~m^{-2}}$. The corresponding mean value $\mu_{\rm ISM}\sim +7.6$ ${\rm rad~m^{-2}}$ is slightly larger than the observed value of $6.3\pm 0.5$ ${\rm rad~m^{-2}}$ \citep{mao10}, but this small difference does not affect our results.

\subsection{RM Associated with Intervening Galaxies (EXG)}
\label{subsection2.4}

External galaxies can intervene along the LOS. If we define $\sigma_{\rm EXG} = (\sigma_{\rm RM}^2-\sigma_{\rm INT}^2-\sigma_{\rm IGM}^2-\sigma_{\rm ISM}^2-\sigma_{\rm ERR}^2)^{1/2}$ and adopt $\sigma_{\rm ERR}= 10.0$ ${\rm rad~m^{-2}}$, $\sigma_{\rm EXG}$ is $4.8-18.4$ ${\rm rad~m^{-2}}$, as shown by the magenta stars in Figure~\ref{f1}. It is, however, difficult to model $\sigma_{\rm EXG}$, since its properties are almost unknown.

Depolarization may be the diagnostic that can be used to identify sightlines containing intervening galaxies. Since the RM due to the IGMF is expected to have scales of $\sim 0.1-1$ degree \citep{ar11} and thus a gradient of the RM within a beam of $\lesssim 10$ arcsec is small, beam depolarization would not take place. On the other hand, RM structure in external galaxies is generally smaller than the beam size, which causes the observed RM and fractional polarization to vary with the observing wevelength, as a result of depolarization \citep{ber12}. Assuming a standard deviation of $\sim 10$~${\rm rad~m^{-2}}$ in intervening galaxies, such depolarization can be seen in the frequency range $\sim 700-1800$ MHz to be covered with the ASKAP and SKA \citep{ab11}.

An absence of such depolarization signals could be a powerful way of identifying intervening galaxies. Therefore, instead of considering RMs of intervening galaxies, we look for sources toward which intervening galaxies do not occur; we refer to such cases as ``no-EXG'' sources. Let us evaluate the chance, $f_{\rm c}$, of an encounter with an intervening galaxy along a LOS. For simplicity, we suppose that the projected surface area of a galaxy is $\sim (30~{\rm kpc})^2$ and that there are $1-10$ galaxies per $(1~{\rm Mpc})^3$. The total number of galaxies in a filament of volume $(10~{\rm Mpc})^3$ is thus $10^{3}-10^{4}$, and the surface filling factor of each galaxy is $\sim (30~{\rm kpc})^2/(10~{\rm Mpc})^2 = 10^{-5}$. A LOS toward a distant radio source ($z>1$) passes through about ten filaments, since the total path-length across typical filaments with IGM temperature $10^5-10^7~{\rm K}$ is $\sim 100~{\rm Mpc}$ \citep[see Figure 6 of][]{ar11}. Therefore, neglecting the overlap of galaxies within the FOV, we estimate $f_{\rm c}\sim (10^{3}-10^{4}) \times 10^{-5} \times 10 \sim 0.1-1$.

Small fractional polarization and its correlation with RRM could be an indicator of depolarization. Actually, fractional polarization correlates with RRM for sources with relatively small fractional polarization in the Hammond catalog. Otherwise, for sources with relatively large fractional polarization, fractional polarization is almost independent of RRM. If we choose sources that have fractional polarization larger than 4 \% as an example of the criterion, we obtain 1776 out of 3650 sources ($f_{\rm c} \sim 0.49$). Note that the 1776 sources are distributed broadly in redshift, suggesting that the source selection would not selectively exclude high-redshift sources.

It has also been argued that RMs of intervening galaxies correlate with optical absorption-line systems \citep{ber08}. Hence, no-EXG sources may be also identifiable via optical spectroscopy. Optical absorption-line data to high-redshift ($z>2$) objects are already available \citep{zhu13}, and in future we expect that data will be obtained toward more sources. So far, \cite{zhu13} have found 40,429 Mg\,{\sc ii}\ absorbers in the spectra of 84,534 quasars in the Sloan Digital Sky Survey. This indicates $f_{\rm c} \lesssim 0.5$, since some SDSS QSOs have multiple absorbers. 

Based on the above results, we adopt $f_{\rm c}=0.5$ as a conservative value. We suppose that half of sources are no-EXG sources and that they are located randomly in the FOV. For no-EXG sources, we adopt $\mu_{\rm EXG}=\sigma_{\rm EXG}= 0$ ${\rm rad~m^{-2}}$.

\subsection{RM due to Observational Errors (ERR)}
\label{subsection2.5}

Other possible uncertainties in RM, such as instrumental noise, calibration error and ionospheric contamination, are modelled as $\mu_{\rm ERR}$ and $\sigma_{\rm ERR}$. Although the errors in existing data are relatively large --- $8~{\rm rad~m^{-2}}$ for VLA RMs \citep{sts11} and $3-5~{\rm rad~m^{-2}}$ for ATCA and WSRT data \citep{mao10} --- this will improve substantially in forthcoming observations. For instance, \cite{sb13} recently reported calibration of ionospheric RMs with absolute errors $\lesssim$ 0.1 ${\rm rad~m^{-2}}$. In this paper, we consider Gaussian errors with $\mu_{\rm ERR}=0$ ${\rm rad~m^{-2}}$ and $\sigma_{\rm ERR}=$ 1, 3, or 5 ${\rm rad~m^{-2}}$ as reachable values for the SKA and its precursors. We focus on results for the case of $\sigma_{\rm ERR}=$ 1 ${\rm rad~m^{-2}}$. Results with $\sigma_{\rm ERR}=$ 3 and 5 ${\rm rad~m^{-2}}$ are presented in Section 5.

\section{Calculation}
\label{section3}

We describe below how we construct two dimensional RM maps and how we perform statistical analysis.

Two-dimensional RM maps are constructed as follows. We consider a FOV of $30^\circ \times 30^\circ$ toward the south Galactic pole, and randomly distribute polarized sources over the distribution in redshift adopted in \S\ref{subsection2.1}. The FOV consists of $16384 \times 16384$ pixels divided evenly; one source is placed in each pixel, and we use the coordinate of the pixel as the position of that source. We explore structures down to $\sim 0.01^\circ$ scales, for which the minimum separation of sources in the FOV, $\sim 0.0018^\circ \sim 0.1'$, is sufficient for our study.

We define the total number of sources as $D \times (30^\circ \times 30^\circ)$, and study cases with source densities $D=1-1000$ ${\rm deg^{-2}}$. Here, it is not essential for our study to specify which observing project is considered, since our demonstrations are made for a given RM grid rather than for a given facility. Source counts to be obtained with future telescopes such as the SKA and ASKAP are a topic of current discussion. Recent works provide careful estimates of source counts \citep{hal14a,hal14b,rud14,sti14}. A summary of some of these estimates can be seen in Figure 4 of \cite{mao14}, which indicates that source density that we have adopted is reasonable, given the overall uncertainty of a factor of a few in current source density estimates.

In our analysis, we do not use sources whose LOSs go through galaxy clusters, based on the X-ray criteria (Section 2.2). In practice, sources behind known clusters can be excluded \citep[e.g. Coma cluster,][]{mao10}. For the remaining sources, we calculate maps of INT, IGM, ISM and ERR, then build the COM map by summing these components. Statistics as presented below are then calculated using the cluster-subtracted sources. From the COM map, we try to extract the IGM map by filtering sources. We study the extent to which statistics for the filtered sources match statistics for the cluster-subtracted sources in the IGM map.

The filtering process is as follows. First, we discard 50 \% of sources for which the LOS passes through intervening galaxies. No-EXG sources can be identified in real data with depolarization and/or optical counterparts (Section 2.4). Second, we exclude nearby sources ($z_{\rm s}<z_{\rm c}$) to reduce the contribution of intrinsic RMs, where $z_{\rm c}$ is a threshold redshift and we assume that we know the redshifts of all detected radio sources. We define $D'$ ${\rm deg^{-2}}$ as the sky density of the remaining, filtered sources. Finally, we subtract the large-angular-scale structure mostly induced by the ISM map by applying a high-pass filter. The residual RM (RRM) map then corresponds to our estimate of the IGM map. The RRM is given by
\begin{equation}\label{eq:RRM}
{\rm RRM}(ix,iy)={\rm COM}(ix,iy) - {\rm MRM}(ix,iy)~,
\end{equation}
where $(ix,iy)$ are the coordinates in the map, and MRM is the mean RM obtained by averaging RMs of the filtered sources over a smoothing diameter, $\theta_{\rm c}$ centered at $(ix,iy)$. In the calculation of MRM, we exclude the source we are trying to filter from the count, and iteratively exclude aberrant sources, defined as sources with an RM more than $3\sigma$ from the mean.

\begin{figure}[tp]
\includegraphics[width=\linewidth]{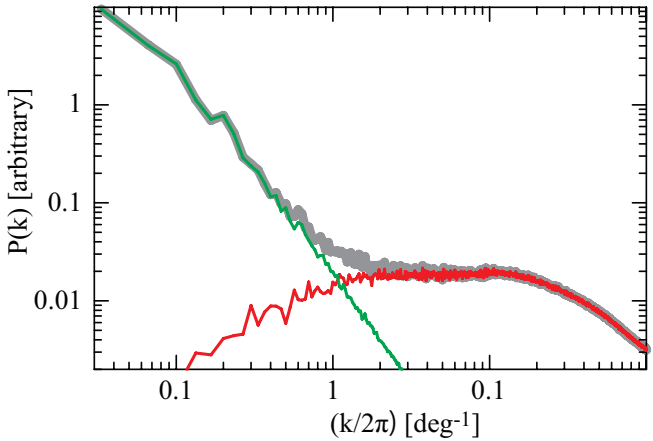}
\caption{
Power spectra of RM maps calculated with $16384 \times 16384$ sources for a $30^\circ \times 30^\circ$ FOV. The red, blue, and gray lines show power spectra for the IGM, ISM, and IGM+ISM, respectively. \label{f2}
}
\end{figure}

\placefigure{f2}

Figure~\ref{f2} shows the power spectra of the IGM and ISM maps. We see that the power of the IGM map overtakes that of the ISM map at $\sim 1-2^\circ$, suggesting $\theta_{\rm c}\sim 1^\circ-2^\circ$ as a good choice for the high-pass filter. A small contribution from the ISM may remain in the RRM map, but the contribution should be several times smaller than the IGM.

Another consideration for deciding $\theta_{\rm c}$ is that sufficient neighboring sources should exist within a diameter $\theta_{\rm c}$ to ensure a reasonable foreground removal. We define $N$ as the number of neighboring sources within $\theta_{\rm c}$, and decided to exclude sources with $N\le 3$ from our statistical analysis. The mean number of neighboring sources is given by $\pi(\theta_{\rm c}/2)^2D'$, i.e., $0.79D'$, $3.14D'$ and $19.6D'$ for $\theta_{c}=$ $1.0^\circ$, $2.0^\circ$ and $5.0^\circ$, respectively. Therefore, to satisfy $N>3$, we adopt $\theta_{\rm c}=5^\circ$ if $D'<1$~deg$^{-2}$, and $\theta_{\rm c}=2^\circ$ for other values of $D'$. As a result of these choices, most of sources have $N\gg 3$, and our results do not dramatically change if we allow $N>2$ or if we include the source we are trying to filter in the count.

It should be noted that we can utilize a larger number of sources in the calculation of RRM, if we apply a high-pass filter before filtering sources. In this case, however, we confirmed that the resultant RRM map substantially underestimates the IGM map, since a corresponding MRM map always overestimates the ISM map due to the contribution of intrinsic RMs of low-redshift sources. Therefore, the high-pass filter should be applied after filtering sources. The same argument is available for RMs due to intervening galaxies.

\section{Results}
\label{section4}

\subsection{Probability Distribution and Standard Deviation}
\label{section4.1}

\begin{figure}[tp] 
\includegraphics[width=\linewidth]{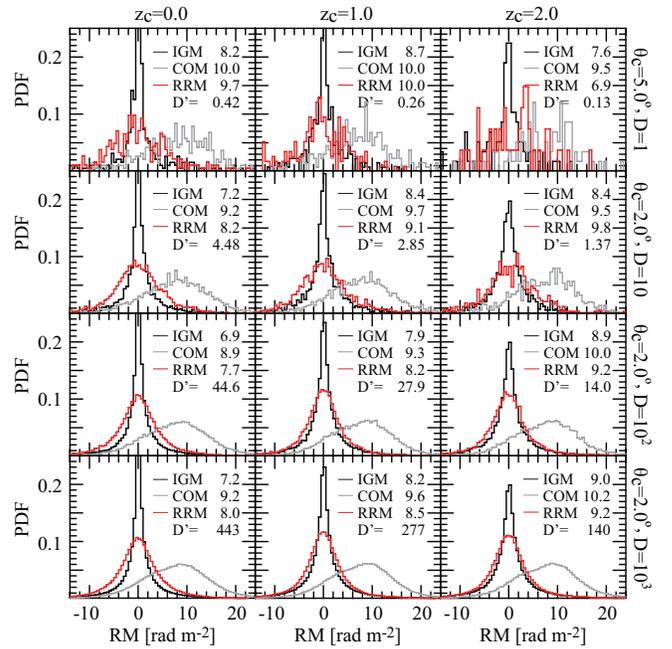} 
\caption{
Probability distribution functions (PDFs) of RM maps. A FOV with $30^\circ \times 30^\circ$ toward the south Galactic pole is considered. Panels from left to right show the results for threshold redshifts, $z_{\rm c}=$ 0.0, 1.0, 2.0, respectively. Panels from top to bottom show the results for source densities, $D=$ 1, 10, $10^2$, and $10^3$~deg$^{-2}$, respectively, and the density of used sources ($D'$) is given in each panel. Smoothing diameters, $\theta_{\rm c}$, are $5.0^\circ$ for $D=1$~deg$^{-2}$ and $2.0^\circ$ for the others. Black, gray, and red lines show the PDFs for the IGM, COM and RRM maps, respectively. Values following the names of components in each panel indicate the standard deviations of RM for the components in rad~m$^{-2}$. \label{f3}
}
\end{figure}

\placefigure{f3}

We first calculate the probability distribution function (PDF) and the standard deviation of RM to derive statistical properties of the modelled and filtered RM maps. Figure~\ref{f3} shows the results for the IGM (black), COM (gray), and RRM (red) maps. Panels from left to right show the results for different threshold redshifts, and panels from top to bottom show the results for different source densities.

Overall, the PDF for $D=1$~deg$^{-2}$ (corresponding to current observational capabilities) has large statistical uncertainties due to a lack of usable sources ($D'=$ 0.42, 0.26, and 0.13~deg$^{-2}$ for $z_{\rm c}=$ 0.0, 1.0, and 2.0, respectively). But for all cases, even with $D=1$~deg$^{-2}$, we confirmed that the non-zero mean of $+7.6~{\rm rad~m^{-2}}$ seen in the COM map (gray) is satisfactorily removed in the RRM map (red). In other words, the high-pass filter is effective in removing large-scale ($\ga$ several degree) coherent structures of RM caused mostly by the ISM map, even for currently obtainable source densities. Statistical uncertainties are greatly improved for $D=10$~deg$^{-2}$ and become almost negligible for $D\ge 100$~deg$^{-2}$. The high-pass filter becomes more accurate and adequately removes structures down to a scale of a few degrees for $D\ge 10$~deg$^{-2}$.

We see that the RRM map still contains a large RM variance after we removed the large-scale structure in RM due to the ISM (Figure~\ref{f2}). The variance cannot be ascribed to EXG RMs, since we have already removed the sources that may have large EXG RMs. Therefore, if ERR RMs (we input $\sigma_{\rm ERR}=$ 1 ${\rm rad~m^{-2}}$) are sufficiently small compared to the standard deviation of $\sim 7-10$~${\rm rad~m^{-2}}$ for the RRM map, the variance in the RRM map can be mostly ascribed to INT and IGM RMs.

Let us now use only high-redshift sources for which INT RMs should be small ($\sigma_{\rm INT}\lesssim 1-2~{\rm rad~m^{-2}}$) in our model. The results with $z_{\rm c}=$ 1.0 or 2.0 clearly indicate that the RRM map still has significant levels of RM variance, which can be attributed to IGM RMs. We find that the standard deviation of the RRM map for high-redshift sources nicely reproduces that of the input IGM map. For instance, relative differences of $\sigma_{\rm RRM}$ to $\sigma_{\rm IGM}$ are $\sim 8-17~\%$ for $D=10$~deg$^{-2}$, $\sim 3-4~\%$ for $D=100$~deg$^{-2}$, and $\sim 2-4~\%$ for $D=1000$~deg$^{-2}$. Note that the standard deviation of the IGM map for distant sources is larger by $\sim 1-2~{\rm rad~m^{-2}}$ than that for all sources, since distant sources tend to have larger RMs (Figure~\ref{f1}).

A good reconstruction can be also seen in the PDF. For sources with $z_{\rm c}=0.0$, the PDF of the RRM map (red) has a broader profile than that of the IGM map (black) due to INT RMs, even for $D=1000$~deg$^{-2}$. Such difference becomes substantially small, if we use only high-redshift sources. Here, the PDF of the RRM map always has a less sharply peaked profile than that of the IGM map, because of ERR RMs of $\sigma_{\rm ERR}=1~{\rm rad~m^{-2}}$.

We have also investigated the cases with $\theta_{\rm c}=1^\circ$ and $5^\circ$ for $D\ge 10$~deg$^{-2}$. We find that for $\theta_{\rm c}=5^\circ$ (not shown in Figure~3), we obtain a broader PDF and a larger ($\sim$sub ${\rm rad~m^{-2}}$) standard deviation than when using $\theta_{\rm c}=2^\circ$.  This is because, when we increase the smoothing diameter, the RRM map contains larger-scale components mostly induced by the ISM map. Such components become a source of error in reconstruction of the IGM map. The best choice of value for $\theta_{\rm c}$ depends on the actual structure in RM for the IGM and ISM maps. The optimal value is $\theta_{\rm c}\sim 1^\circ-2^\circ$ for our models, since the power of the IGM map overtakes that of the ISM map (Figure~\ref{f2}) at these scales. The case for $\theta_{\rm c}=1^\circ$ for $D\ge 10$~deg$^{-2}$ gives similar results to those shown for $\theta_{\rm c}=2^\circ$ in Figure~3.

\subsection{Second-Order Structure Function}
\label{section4.2}

\begin{figure}[tp]
\includegraphics[width=\linewidth]{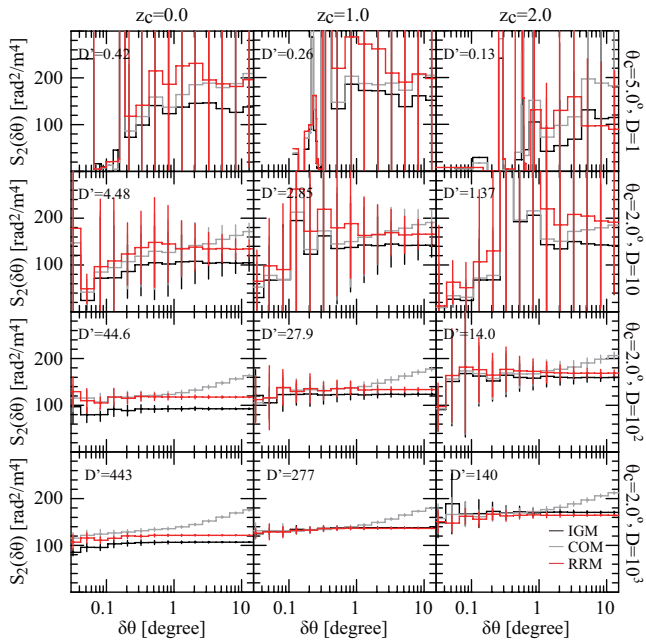}
\caption{
Second-order structure functions (SFs) of RM maps. A FOV with $30^\circ \times 30^\circ$ toward the south Galactic pole is considered. Parameters for each panel are the same as in Figure~\ref{f3}. Black, gray, and red lines show the SFs of IGM, COM, and RRM, respectively. $D'$ is the density of usable sources. The SF is calculated with a spatial resolution of $0.0018^\circ$, then binned in equal log intervals of 0.2. For each bin, we calculate the average and the standard deviation, which are drawn as lines and error bars, respectively. \label{f4}
}
\end{figure}

\placefigure{f4}

We next calculate the structure function (SF) of RM. The SF tells us at which angular scales the spatial structure of RM decorrelates. We also considered the power spectrum of RM, but we did not obtain reasonable results since the Fourier transform of unevenly-sampled data generated huge numerical errors. It is not obvious how one can treat blank pixels in the map, and we did not remove such errors by simple interpolation of the data. Further sophisticated procedures are thus needed to derive meaningful power spectra.

The $n$-th order SF is defined as
\begin{equation}\label{eq:SF}
S_n(\delta\theta)=\langle|RM(\theta+\delta\theta)-RM(\theta)|^n\rangle_\theta~,
\end{equation}
where the subscript $\theta$ on the right-hand side indicates averaging over the data domain of $\theta$. We calculate the second-order SF, $S_2$, at a spatial resolution of $0.0018^\circ$. We then bin the SF in equal log intervals in the same manner adopted in previous works \citep{mao10,sts11}. We adopt a log interval of 0.2 and for each bin we calculate the average and standard deviation of the SF. Results for different values of $D$ and $z_{\rm c}$ are shown in Figure \ref{f4}, where the average of the SF within each bin and the standard deviation (the scatter) of the SF within each bin are drawn as lines and error bars, respectively.

Overall, $S_2$ for $D=1$~deg$^{-2}$ corresponding to current observational capabilities has large statistical uncertainties due to a lack of usable sources. We require $D'\ga 4$~deg$^{-2}$ to obtain $S_2$ down to scales of $\sim 1^\circ$ with errors less than $30\%$, and such accuracy is marginally achieved for $z_{\rm c}=0.0$ with $D=10$~deg$^{-2}$. Uncertainties are greatly improved for the cases with $D=100$~deg$^{-2}$ for which we obtain clear $S_2$ on scales down to sub degrees. For $D=1000$~deg$^{-2}$ as achievable with the SKA, $S_2$ could be studied even down to scales of $\sim 0.1^\circ$ with errors less than $\sim 30~\%$.

Looking at cases with sufficient source densities (e.g. $D\ge100$~deg$^{-2}$), we see that $S_2$ for the RRM map (red) has a flat profile at scales $\delta\theta\ga 1^\circ$. This is also evidence that the high-pass filter has satisfactorily removed most ISM RMs, which show a monotonic increase of $S_2$ from $\sim 0.01^\circ$ to $10^\circ$ \citep{arkg13} as can be partly seen in $S_2$ for the COM map (gray). Recall that we have already removed the sources that may have large EXG RMs; $S_2$ for the RRM map is thus ascribed to INT, IGM, and ERR RMs. Here, $S_2$ for the INT and ERR maps has a flat profile over $\sim 0.01^\circ-10^\circ$ (not shown), because these structure functions have white-noise spectra in Fourier space. Therefore, these contributions enhance $S_2$ at all scales shown, and the amplitude of $S_2$ for the RRM map (red) becomes somewhat larger than that for the IGM map (black). Such white-noise power is difficult to selectively remove with structure-based filters. Instead, we can use only high-redshift sources for which INT RMs should be small in our model. The results with $z_{\rm c}=$ 1.0 or 2.0 clearly indicate that $S_2$ for the IGM map is successfully reconstructed, if we reach $D=1000$~deg$^{-2}$ corresponding to SKA observations. There remains the contribution of ERR RMs, but this is not problematic for the study of the SF, provided that $\sigma_{\rm ERR}=1~{\rm rad~m^{-2}}$ or less.

An important feature of the SF for the IGM predicted by \cite{ar11} is a decline at smaller scales, $\delta\theta\lesssim 0.1^\circ$. Such a decline is hidden by uncertainties for the cases with $D\la 100$~deg$^{-2}$, but might be seen with $D'\gtrsim 300$~deg$^{-2}$. Interestingly, although the SF of the RRM map for $z_{\rm c}=0.0$ overestimates that of the IGM map, the decline can be seen for $D=1000$~deg$^{-2}$ with the SKA, regardless of the redshift criterion $z_{\rm c}$. This is because the IGM map is a major component in the COM map at these scales.

We have also investigated the cases with $\theta_{\rm c}=1^\circ$ and $5^\circ$ for $D\ga 10$~deg$^{-2}$. We find that for $\theta_{\rm c}=5^\circ$ (not shown), the SF of the RRM map shows a slight increase at $\theta \gtrsim 1^\circ$, induced by the ISM map (not shown). This behavior is consistent with that seen in the PDF (Figure~\ref{f3}); the RRM map contains larger-scale components as we increase the smoothing diameter, mostly induced by the ISM map. For $D\ge 10$~deg$^{-2}$, the results for $\theta_{\rm c}=1^\circ$ are similar to those for $\theta_{\rm c}=2^\circ$.

\section{Discussion}
\label{section5}

\subsection{Other Scenarios}
\label{section5.1}

We have presented results for the most probable scenario in which there are RM contributions from the source, the IGM, intervening galaxies, the ISM, and observational errors. Although the scenario is based on observations and simulations, one may consider alternatives. Particularly, it would be insightful to consider the case in which the RM of filaments is insignificant. Therefore, in this subsection we consider other models for RM contributions. We expect that the models themselves will be improved with future observations.

\subsubsection{Large INT instead of IGM}
\label{section5.1.1}

\begin{figure}[tp]
\includegraphics[width=\linewidth]{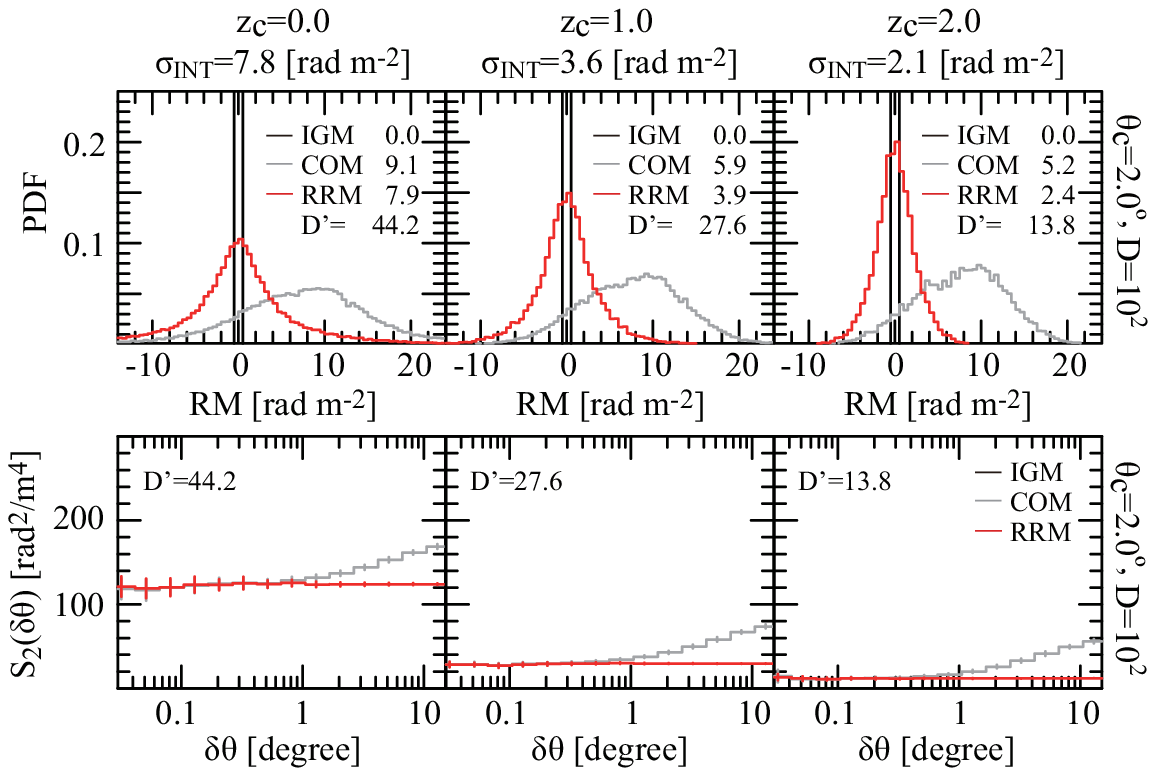}
\caption{
As for Figure \ref{f3} (top panels) and Figure \ref{f4} (bottom panels) for $D=100$~deg$^{-2}$ and $\theta_{\rm c}=2.0^\circ$, but with $\sigma_{\rm IGM}=0.01$ ${\rm rad~m^{-2}}$ and $\sigma_{\rm INT}=7.8$ ${\rm rad~m^{-2}}$.\label{f5}
}
\end{figure}

\placefigure{f5}

First, we vary INT RMs. As discussed in \S\ref{subsection2.1}, there is a possible range $\sigma_{{\rm INT},0}\sim 7-13$ ${\rm rad~m^{-2}}$. If we adopt $15~{\rm rad~m^{-2}}$, we find $\sigma_{\rm INT}=\sigma_{{\rm INT},0}(1+z)^{-2}\sim 1.7$ ${\rm rad~m^{-2}}$ for sources with $z>2$, which is still insignificant and does not change our main results. If the RM due to the IGMF is close to zero and the intrinsic RM dominates the observed RMs, the intrinsic RM is required to be $\sim 2.55$ times larger than the value we adopted. The results of this case are shown in Figure~\ref{f5}. In this case, we would clearly see a $(1+z)^{-2}$ relation in the standard deviation and the PDF of the RRM map. The SF of the RRM map would also decrease with increasing $z_{\rm s}$, showing a flat profile.

Such a strong dependence of RM on redshift is, however, not observed \citep{ham12}, and the resultant standard deviation of $22.5~{\rm rad~m^{-2}}$ for $z\sim 0$ is too large compared with the possible range described above. These results suggest that if intrinsic RMs dominate the observed RMs, the standard deviation of RM should follow $\sigma_{\rm INT}=\sigma_{{\rm INT},z}(1+z)^{-2}$ and $\sigma_{{\rm INT},z}\propto (1+z)^{n}$ with $n>0$, i.e., sources at higher-redshifts will have larger RMs. Note that observed intrinsic RMs are a result of competition between source evolution and depolarization. The actual dependence will be related to the effects of active galactic nuclei and star-formation, the masses of galaxies, and the bias of observations. Since in our approach the redshift dependence of $\sigma_{{\rm INT},z}$ is essential for reducing the contribution of INT RMs from observed RMs, further studies of radio sources, e.g. ultra-high resolution observations with SKA, are of crucial importance.

\subsubsection{Large ISM instead of IGM}
\label{section5.1.2}

\begin{figure}[tp]
\includegraphics[width=\linewidth]{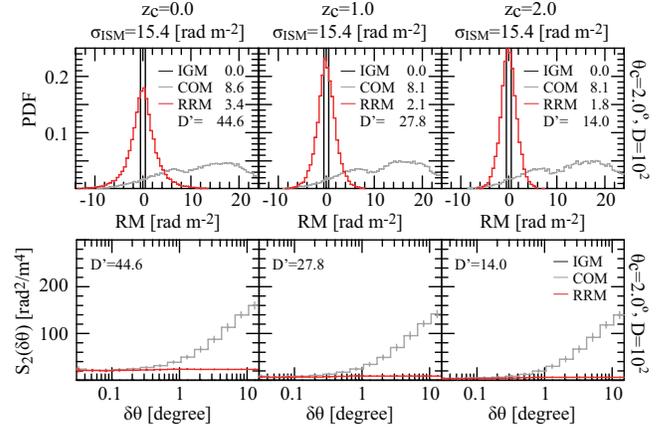}
\caption{
As for Figure \ref{f3} (top panels) and Figure \ref{f4} (bottom panels) for $D=100$~deg$^{-2}$ and $\theta_{\rm c}=2.0^\circ$, but with $\sigma_{\rm IGM}=0.01$ ${\rm rad~m^{-2}}$ and $\sigma_{\rm GMF}=15.4$ ${\rm rad~m^{-2}}$. \label{f6}
}
\end{figure}

\placefigure{f6}

We could also consider the case of a large Galactic RM dominating the observed Faraday rotation. In this case, the Galactic RM is required to be $\sim 1.73$ times larger than that adopted in \S\ref{subsection2.3} to explain the standard deviation of the observed RMs. Since the Galactic contribution is mostly removed by a high-pass filter, in this case we obtain a very small standard deviation and a narrow PDF for the RRM map (Figure~\ref{f6}), both of which show a $(1+z)^{-2}$ relation caused by the INT map. The SF of the COM map will go down to $O(10)$ ${\rm rad~m^{-2}}$ at a scale of $\sim 1^\circ$. Such a decline is not observed \citep{mao10,sts11}, although present observations still have large uncertainties at sub degree scales. The SF at sub degree scales could begin to be studied with the ASKAP. Note that the estimation of the RM due to the IGMF would be easier if we were to consider the north Galactic pole, since the average and the standard deviation are then both smaller \citep{arkg13}. 

A more critical change to the ISM model would be to increase the power at small scales. If the ISM component has significant RMs at scales less than $\sim 1^\circ$, a high-pass filter will fail to remove the ISM component from the RRM map. Actually, it has been suggested that there are small-scale structures in RM in the Galactic plane \citep[e.g.,][]{hav06,hav08}. We have adopted the Milky Way model of \cite{arkg13}, which is based on observed properties of turbulence. The model has incorporated small-scale structures caused by turbulence, and the power of which in small scales ($\lesssim 1^\circ$) is negligibly small toward high Galactic latitudes. Future observations will provide denser RM grids, and may allow to study small-scale Galactic RM fluctuations not originating from turbulence toward high Galactic latitudes.

Recently, a new method using millisecond pulsars in globular clusters has been proposed to study small-scale ISM magnetic fields \citep{ho14}. \cite{ho14} have inferred 0.1 $\mu$G fluctuations on parsec scales toward high Galactic latitudes, which is about one order of magnitude weaker than the strength of turbulent magnetic fields adopted by \cite{arkg13}. This may be evidence that sub pc-scale Galactic magnetic fields are not predominant, but does not dramatically alter the power spectrum and the structure function of RMs in our model.

\subsubsection{Large ERR instead of IGM}
\label{section5.1.3}

\begin{figure}[tp]
\includegraphics[width=\linewidth]{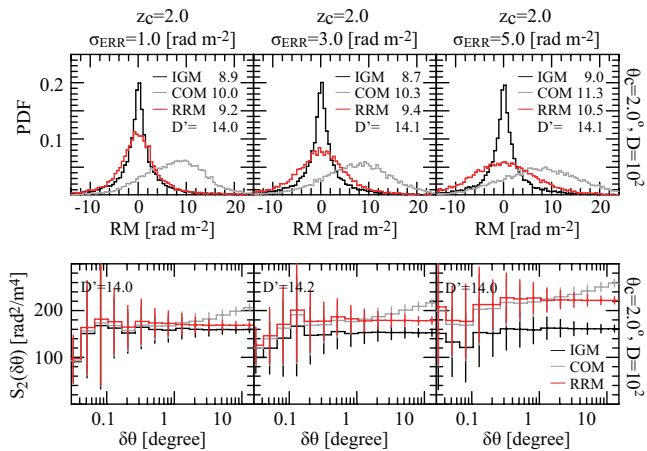}
\caption{
As for Figure \ref{f3} (top panels) and Figure \ref{f4} (bottom panels) for $D=100$~deg$^{-2}$, $z_{\rm c}=2.0$ and $\theta_{\rm c}=2.0^\circ$, but with $\sigma_{\rm ERR}=$, 1.0, 3.0, and 5.0 ${\rm rad~m^{-2}}$ from left to right, respectively. \label{f7}
}
\end{figure}

\placefigure{f7}

The measured properties of the SF of RM depends on the reliability of the source count and error estimates. We have assumed that the source count and error estimates obtained from observations are reliable. Discussion about the reliability of these measurements is beyond the scope of our work, nevertheless, these effects can be considered by showing cases for large errors. Figure~\ref{f7} shows the cases for $\sigma_{\rm ERR}=$, 1.0, 3.0, and 5.0 ${\rm rad~m^{-2}}$. Since the ERR map has a white-noise spectrum in Fourier space, it cannot be fully removed by a high-pass filter. As a result, we see a broader PDF and a larger standard deviation of the RRM map compared with the IGM map. 

Errors change the scales at which intrinsic and Galactic contributions dominate. Generally speaking, the apparent scales of intrinsic and Galactic contributions could change if the error RM is comparable to the observed RM itself. An instructive example can be seen in Figure 5 of \cite{sts11} in which they demonstrated how errors with powers of 0.4-8 times the data values change the slope of the SF. Nevertheless, we expect that errors will become sufficiently small compared to intrinsic and Galactic contributions in future observations.

A possible way to reduce the error term would be a further selection of sources that have relatively small errors, although this will reduce the number of sources and will increase statistical uncertainties. The fraction of the sources that satisfy a given level of RM errors is not well-understood. Further study of this fraction will be helpful to identify a sweet spot between the number of sources and the amplitude of errors.

\subsection{Implication to the IGMF}
\label{section5.2}

Finally, we consider how our results can probe the possible nature of the IGMF. First of all, we have adopted a model of the IGMF that has theoretical uncertainties of up to a factor of a few both on the strength and coherence length \citep{ar10}. There are also IGMF models that we have not considered (see \cite{ar11} and references therein), which could have different strengths and different coherence lengths of the IGMF in filaments compared to that used here.

If we suppose that an RRM map derived according to the approach in this paper can meaningfully recover the average, standard deviation and SF of the IGM map from observed data, the nature of the IGMF can be constrained as follows. Since recent cosmological simulations share a broad agreement on the density structure of the IGM, we can evaluate an electron density and depth of a filament, and the number of filaments along the LOS. Thus, the standard deviation of the RRM map jointly constrains the strength and coherence length of the IGMF \citep[e.g., using Eq. (11) of][]{cr09}. Models predicting peculiarly small or large IGMF strengths could be ruled out from such constraints.

The degeneracy between the strength and the coherence length of the IGMF in RM could be broken using the SF. For instance, if the SF starts to decrease around an angular separation of $0.1^\circ$, this implies a coherence length of the IGMF in filaments of several hundred kpc \citep{ar11}. In contrast, an IGMF with a larger coherence length should show a decline of the SF at scales larger than $\sim 0.1^\circ$, and such a decline could be detected with ASKAP ($D\sim 100$~deg$^{-2}$). But if the IGMF is much weaker and/or the coherence length is much shorter than several hundred kpc, we would see a flat SF at $\sim 0.1^\circ$, and such behavior could only be studied with the SKA ($D\gtrsim 1000$~deg$^{-2}$).

Note that if the Local Group IGMF has a significant RM, this would have uniform distribution within the considered FOV since the angular size of the corresponding coherence length would be $O(10)$ degrees \citep{ar11}. Hence the RM of the Local Group should mainly contribute to the average observed RM, and a high-pass filter will reduce it in the RRM map. To study the RM of the Local Group, we thus need to consider a much wider FOV and need to investigate very large-scale coherent RM structures.

Finally, if the RRM map truly reproduces the IGM map, reconstructed RM data should show a monotonic increase as a function of $z_{\rm s}$ and should show a saturation for large $z_{\rm s}$ (Figure~\ref{f1}). In addition, since the IGM map traces the large-scale distribution of matter \citep{ar10}, reconstructed RMs should have a correlation with tracers of the large-scale structure such as the number density of galaxies, the X-ray surface brightness of the IGM and the Sunyaev-Zel'dovich effect against the cosmic microwave background. Such correlations would confirm the discovery of Faraday rotation due to the IGMF in the cosmic web, and will be considered in future studies.

\section{Conclusion}
\label{section6}

In this paper we have demonstrated an approach for estimating the Faraday rotation measure (RM) produced by the intergalactic magnetic field (IGMF), in which we have incorporated models of RM for polarized sources, the IGM, intervening galaxies, the ISM and observational errors. We have adopted a scenario in which the observer-flame RM of sources decreases with redshift by $1/(1+z)^2$ and the observer-flame RM of the IGMF through filaments accumulates with redshift by the manner predicted from cosmological simulations (Figure \ref{f1}). We considered a $30^\circ\times 30^\circ$ field-of-view toward the Galactic caps, motivated by previous observational studies. We found that a high-pass filter is quite effective at removing the Galactic contribution from the observed RMs. Reductions of RMs $\sim 4.8 - 18.4$ rad m$^{-2}$ caused by intervening galaxies and observational errors more than a few rad m$^{-2}$ are both critical for the study of the RM due to the IGMF. After selecting a half of observed sources toward which sightlines do not contain intervening galaxies, and assuming that RM errors are less than a few rad m$^{-2}$, our approach allows us to estimate the standard deviation of the RM due to the IGMF with errors less than $\sim 20~\%$ and $\sim 4~\%$ if source densities per square degree of $D\sim 10$ and $D\ge 100$~deg$^{-2}$ are available, respectively. The second-order structure function of the RM due to the IGMF will be able to be studied with errors less than $\sim 30~\%$ down to scales of $\sim 0.1^\circ$ if we can achieve a sky density $D\sim 1000$~deg$^{-2}$ for polarized extragalactic radio sources in the SKA era.

\acknowledgments

T.A. is supported by the Japan Society for the Promotion of Science (JSPS). B.M.G. is supported by the Australian Research Council through grant FL100100114. D.R. acknowledges the support of the National Research Foundation of Korea through grant 2007-0093860.


\end{document}